\pdfoutput=1
\documentclass[11pt,superscriptaddress]{revtex4}
\usepackage[pdftex]{graphicx}
\usepackage{amsmath}
\usepackage{amssymb}
\usepackage{amsxtra}

\newcommand\half{\textstyle{\frac{1}{2}}}
\newcommand{\MeV}{\,\mathrm{Me\kern -1pt V}}       

\begin{document}
\title{Are superheavy stable quark clusters viable candidates for the dark matter?}
\author{Norma Manko\v{c} Bor\v{s}tnik}
\affiliation{Faculty of Mathematics and Physics, University of Ljubljana,
     Jadranska~19, P.O.~Box 2964, 1001 Ljubljana, Slovenia}
\author{Mitja Rosina}
\affiliation{Faculty of Mathematics and Physics, University of Ljubljana,
     Jadranska~19, P.O.~Box 2964, 1001 Ljubljana, Slovenia}
\affiliation{J.~Stefan Institute,  1000 Ljubljana, Slovenia}
\begin{abstract}.   
The explanation for the origin of families of quarks and leptons and their properties is 
one of the most promising ways to understand the assumptions of the {\it Standard Model}. 
The {\it Spin-Charge-Family} theory~\cite{M92,M93,M01,M95,M1315,matterantimatter,scalar,%
AM,BM,BBLM,GN2014,Bled}, which 
does propose the mechanism for the appearance of families and offers an explanation for all 
the assumptions of the {\it Standard Model}, predicts two decoupled groups of four families.  
The lightest of the upper four families has stable members, which are correspondingly 
candidates to constitute the dark matter~\cite{BM}. We study the weak and the "nuclear" 
(determined by the colour interaction among the heavy fifth family quarks) scattering  of such 
a very heavy baryon 
by ordinary nucleons in order to show that the cross-section is very small and consistent with the
observation in most experiments so far, provided that the quark mass of this baryon is about 
100 TeV or above. 
\end{abstract}
\maketitle 

\section{Introduction}
\label{introduction}

The {\it Standard Model} has no explanation for  the existence of families and their 
properties as well as for the appearance of the scalar field, which in the {\it Standard Model} 
determines the properties of the electroweak gauge fields and, together with the Yukawa 
couplings, the properties of families of quarks and leptons. 
A theory which would explain the origin of families and the mechanism causing the observed 
properties of quarks, leptons and gauge fields is needed. 

The {\it Spin-Charge-Family } theory~\cite{M92,M93,M01,M95,M1315,matterantimatter,scalar,%
AM,BM,BBLM,GN2014,Bled}, which offers the explanation for all the assumptions of the {\it Standard
Model}, with the appearance of families, of the Higgs's scalar and the Yukawa couplings included,
is very promising for this purpose.  
The theory predicts that there are the stable fifth family baryons, which constitute the dark 
matter~\cite{BM}. The main features of this theory are summarized in 
Sect.~\ref{normatheory}.

The properties of the fifth family members, their freezing out in the evolution of the universe 
and also their interaction with the ordinary matter were evaluated in Ref.~\cite{BM}. In this 
paper we study scattering of these stable fifth family baryons on the ordinary nucleons, when 
this very small object of the mass about one hundred TeV approaches with the velocity of our 
Sun an ordinary nucleus. The nucleus, which is six orders of magnitude larger and three orders 
of magnitude lighter than the very heavy baryon, recoils elastically from the very heavy 
baryon, as long
as is the excitation energy of the ordinary nucleus much smaller than the kinetic energy of the  
relative motion of both objects. We test  whether the results agree with the present 
experiments~\cite{DAMA,rita,CDMS,CRESST,XENON100,BML,K}.

In this paper we present:\\
{\bf i.} We show under which conditions the  electrically neutral baryon n$_5$ = 
(u$_5$d$_5$d$_5$) is the lightest~\cite{BM} (the subscript 5 is used to denote the fifth-family 
members).
One might expect that a charged combination (u$_5$u$_5$u$_5$) or (d$_5$d$_5$d$_5$)
would be the lightest, depending on whether u$_5$ or d$_5$ is lighter (what depends on the 
scalar and other interactions, which determe masses of the stabel fifth family members). 
We present limits on the u-d mass difference such that the neutral combination is in fact the 
lightest. \\
{\bf ii.} Since the quark clusters always carry the weak charge, we present the constraint that 
the corresponding weak scattering does not exceed  the upper bound estimated in present
observations. It turns out that in the (light cluster) - (very heavy cluster) scattering 
the weak cross section limits to a value around $10^{-13} \mathrm{fm}^2$. 
This means that the low rate of collisions of the fifth family candidate for the dark matter with 
the ordinary matter requires a low number density of dark matter particles to be in agreement 
with the observations. From the dark matter mass density~\cite{DATA?} it follows that the 
fifth family baryon mass must be high - around 100 TeV or above.\\
{\bf iii.} To binding of the ordinary quarks in nucleon many gluons contribute. The gluon field
around the quarks makes the main contribution to the gluon mass, manifesting 
in the potential which grows with the distance between two quarks. When the ordinary  nucleon 
(of the size of fm) scatters  on another ordinary nucleon, the nuclear force always dominates 
over the weak force. 
Very heavy quarks are bound, due to the heavy mass, mostly through one gluon exchange
force.
When a light ordinary nucleus scatter on a colour neutral very heavy baryon, a first family 
quark of the nucleon sees the colour force of the heavy baryon only due to the induced colour 
dipole - colour dipole, and higher multipole, interaction.
We explore smallness of the colour dipole - colour dipole cross section in dependence of the 
mass of the fifth family quarks. 
We show that  the very heavy quark masses, which satisfy the bound for weak scattering, 
also satisfy the bound for strong scattering.
 
This paper follows to some extent the Ref.~\cite{MN1012,Bled2010,Bled2012}. 

\section{The Spin-Charge-Family theory}
\label{normatheory} 

In this section  a  short introduction to the {\it Spin-Charge-Family} theory~\cite{M92,M93,M01,M95,%
M1315,matterantimatter,scalar,AM,BM,BBLM,GN2014,Bled,BBLM} is presented. Only the  essential 
things arereviewed. The reader can skip this section and continue with the next one, which is the 
main contribution of this paper. We hope, however, that this section might make the curious  reader 
to start thinking about what new - in relation to other proposals - in answering the open problems
 in elementary particle physics and cosmology does this theory bring and, in particular with respect 
to this paper, the reader might start to think about the 
differences in the hadronic properties of the very heavy fifth family hadrons as compared to
the lowest three families hadrons. Due to very strongly bound states the fifth family nuclear 
force differs strongly from the ordinary nuclear force.

The {\it Spin-Charge-Family} theory proposes in $d=(1 + (d-1))$ ($d=14$) dimensions a very simple 
action for spinors  $\psi$ - ${\mathcal L}_{f}$ - which carry two kinds of the spin generators 
($\gamma^a$ and $\tilde{\gamma}^a$ operators), and  for the corresponding gauge fields (vielbeins 
$f^{\alpha}{}_{a}$ and the two kinds of the spin connection fields $\omega_{ab\alpha}$ and 
$\tilde{\omega}_{ab\alpha}$, respectively) - ${\mathcal L}_{g}$ - which is linear in the curvatures 
$R$ and $\tilde{R}$. 
\begin{eqnarray}
S &=& \int \; d^dx \; E\;{\mathcal L}_{f} +  \int \; d^dx \; E\; {\mathcal L}_{g}\,, 
\label{action}
\end{eqnarray}
\begin{eqnarray}
     {\mathcal L}_f &=& \frac{1}{2} (\bar{\psi} \, \gamma^a p_{0a} \psi) + h.c. \nonumber\\
p_{0a }&=& f^{\alpha}{}_a p_{0\alpha}, \quad  p_{0\alpha} =  p_{\alpha}  - 
                     \frac{1}{2}  S^{ab}   \omega_{ab\alpha} - 
                    \frac{1}{2}   \tilde{S}^{ab}   \tilde{\omega}_{ab\alpha}\,,
\label{lspinor}
\end{eqnarray}
\begin{eqnarray}
\label{lgauge}
{\mathcal L}_{g} &= & (\alpha \, R + \tilde{\alpha}  \tilde{R})\,, \nonumber\\
R &=& f^{\alpha [a}f^{\beta b]} \;(\omega_{a b \alpha,\beta} - \omega_{c a \alpha}
\omega^{c}{}_{b \beta})\,,\nonumber\\
\tilde{R} &=& f^{\alpha [a} f^{\beta b]} \;(\tilde{\omega}_{a b \alpha,\beta} - 
\tilde{\omega}_{c a \alpha} \tilde{\omega}^{c}{}_{b \beta})\,.
\end{eqnarray}
Spinors carry  in $d=(1+13)$ dimensions two  kinds of the spin represented by the two kinds
 of the 
Clifford algebra objects~\cite{M92,M93,M01,M95,M1315,matterantimatter,scalar,AM,Bled} 
\begin{eqnarray}
\label{clifford}
&&\{\gamma^a, \gamma^b \}_{+}= 2 \eta^{ab}= \{\tilde{\gamma}^a, \tilde{\gamma}^b \}_{+}\,, \quad 
 \{\gamma^a, \tilde{\gamma}^b \}_{+}=0\,,\nonumber\\
&&S^{ab}= \frac{i}{4} (\gamma^a \gamma^b - \gamma^b \gamma^a)\,,\quad   
\tilde{S}^{ab} = \frac{i}{4} (\tilde{\gamma}^a \tilde{\gamma}^b - \tilde{\gamma}^b \tilde{\gamma}^a)\,, 
\quad \{S^{ab},\tilde{S}^{cd}\}_{-} = 0\,,
\end{eqnarray}
and interact correspondingly with the vielbeins and the two kinds of the spin connection fields.
($f^{\alpha}{}_{a}$ are inverted vielbeins to $e^{a}{}_{\alpha}$ with the properties 
$e^a{}_{\alpha} f^{\alpha}{\!}_b = \delta^a{\!}_b,\; 
e^a{\!}_{\alpha} f^{\beta}{\!}_a = \delta^{\beta}_{\alpha} $, $ E = \det(e^a{\!}_{\alpha}) $. 
Latin indices $a,b,..,m,n,..,s,t,..$ denote a tangent space (a flat index), while Greek indices 
$\alpha, \beta,..,\mu, \nu,.. \sigma,\tau, ..$ denote an Einstein index (a curved index). Letters  
from the beginning of both the alphabets indicate a general index ($a,b,c,..$   and $\alpha,$
$ \beta, \gamma,.. $ ), from the middle of both the alphabets the observed dimensions $0,1,2,3$ 
($m,n,..$ and $\mu,\nu,..$), indices from the bottom of the alphabets indicate the compactified 
dimensions ($s,t,..$ and $\sigma,\tau,..$). We assume the signature $\eta^{ab} = diag\{1,-1,-1,
\cdots,-1\}$. $f^{\alpha [a} f^{\beta b]}= f^{\alpha a} f^{\beta b} - f^{\alpha b} f^{\beta a}$.

One kind of the spin, the Dirac  $\gamma^a$ one, explains the spin ($a=(0,1,2,3)$) and all the 
charges ($a=(5,\dots,14))~$\footnote{The operators $\gamma^a$ are matrices after we make a 
choice of the basis and find their representations. One can find in Refs.~\cite{matterantimatter,scalar} 
a short explanation for such a choice of basis. There also the expressions for all the operators - 
those which operates on spins degrees of freedom and those which operates on charge or family 
quantum numbers - can be found.}, the other - 
$\tilde{\gamma}^a$ - explains the appearance of families. We kindly ask the reader to learn about 
all the details of the above action in Refs.~\cite{matterantimatter,scalar} (arxiv:1412.5866), 
where also the breaks of the starting symmetry, which makes the action~(\ref{action}) to manifest 
in $d=(1+3)$ effectively as the {\it Standard Model} action before the electroweak break, with the 
right handed neutrinos included, are presented. 

The theory explains, why the charges of the left handed quarks and leptons (neutrinos are the regular
members of each family) differ from the charges of the right handed ones. One spinor representation 
of $SO(1,13)$ contains, namely, the {\it Standard Model} quarks and leptons: The left handed
members are weak ($SU(2)_{I}$) charged and $SU(2)_{II}$ (this charge determines the hyper
charge) chargeless, while the right handed members are weak ($SU(2)_{I}$) chargeless and 
$SU(2)_{II}$ charged. Quarks carry the colour ($SU(3)$ $\subset$ of $SO(6)$) charge, leptons are  
colour chargeless. (See table IV and table III in Ref.~\cite{M1315}, respectively.) 

The theory explains also the appearance of families, the generators ($\widetilde{S}^{ab}$) of which 
originate in $\widetilde{SO}(1,13)$. (See table V and table IV in Ref.~\cite{M1315}, respectively.) 
The theory predicts an even number of families, indeed two decoupled groups of four families, 
manifesting $\widetilde{SU}(2)_{I\,SO(1,3)} \times \widetilde{SU}(2)_{I\,SO(4)}$ and 
$\widetilde{SU}(2)_{II\,SO(1,3)}$ $ \times \widetilde{SU}(2)_{II\,SO(4)}$, respectively, where 
($\widetilde{SO}(1,3)$, $\widetilde{SO}(4)$ $\subset$ $\widetilde{SO}(1,7)$ $\subset$ 
$\widetilde{SO}(1,13)$). 
The fourth of the lowest four families will be measured at the LHC, the lowest of the upper four 
families are predicted to form the dark matter~\cite{BM}.

Let us add that there is no evidence so far for the existence of the members of the fourth family. 
Calculations~\cite{DATA?}, all model depenedent, in correlation with the measurements, predict
that the fourth family quark masses can not be smaller than $700$ GeV. The strongest limits 
come from the mesons decay~\cite{Vysotsky}, which are, assuming the  existence of only one scalar
field, again model dependent.  
 
One of the authors of this paper (N.S.M.B.) (together with the coauthor G. Bregar)  analyses the 
properties of the lower four families as predicted by the {\it Spin-Charge-Family} theory in 
Ref.~\cite{GN2014}.   In this theory   
several scalar fields~\cite{GN2014} determine masses of fermions and weak bosons, replacing
 the Higgs's scalar and Yukawa couplings of the {\it Standard Model}. 

One can rewrite the fermion Lagrange density ${\mathcal L}_{f}$ so that a part of it (the first term 
in Eq.~(\ref{lsmyukawa})) manifests the dynamical part of the {\it Standard Model} action for
massless fermions, a part of it (the second term in Eq.~(\ref{lsmyukawa})) represents the interaction 
of fermions with the scalar gauge fields which determine, after they gain nonzero vacuum expectation 
values and cause the electroweak break, mass matrices of the two groups of four families of fermions, 
while a part (the third term in Eq.~(\ref{lsmyukawa})) causes transitions of antileptons into quarks and 
antiquarks into quarks (and back)
\begin{eqnarray}
{\mathcal L}_f &=&  \bar{\psi}\gamma^{m} (p_{m}- \sum_{A,i}\; g^{A}\tau^{Ai} A^{Ai}_{m}) \psi 
+ \nonumber\\
               & &  \{ \sum_{s=7,8}\;  \bar{\psi} \gamma^{s} p_{0s} \; \psi \} 
+ \nonumber\\ 
& & \{ \sum_{t=5,6,9,\dots, 14}\;  \bar{\psi} \gamma^{t} p_{0t} \; \psi \}
\,, \nonumber\\
p_{0s} &=&  p_{s}  - \frac{1}{2}  S^{s' s"} \omega_{s' s" s} - 
                    \frac{1}{2}  \tilde{S}^{ab}   \tilde{\omega}_{ab s}\,,\nonumber\\
p_{0t} &=&  p_{t}  - \frac{1}{2}  S^{t' t"} \omega_{t' t" t} - 
                    \frac{1}{2}  \tilde{S}^{ab}   \tilde{\omega}_{ab t}\,.                    
\label{lsmyukawa}
\end{eqnarray}
Here $ m \in (0,1,2,3)$, $s \in (7,8),\, (s',s") \in (5,6,7,8)$, $(a,b)$ (appearing in $\tilde{S}^{ab}$) 
run within $ (0,1,2,3)$ and $ (5,6,7,8)$, $t \in (5,6,9,\dots,13,14)$, $(t',t") \in  (5,6,7,8)$ and 
$\in (9,10,\dots,14)$. 
The spinor function $\psi$ represents all family members of all the $2^{\frac{7+1}{2}-1}$ families,
which are before the electroweak break massless due to the mass protection caused by the fact that
the left and right handed members carry different weak and hyper charges (table III and table IV in 
Refs.~\cite{M1315}).
The operators $\tau^{Ai}$ ($= \sum_{a,b} \;c^{Ai}{ }_{ab} \; S^{ab}$) determine the hyper 
charge ($A=1$), the weak charge ($A=2$) and the colour charge ($A=3$): $\{\tau^{Ai}, 
\tau^{Bj}\}_- =$ $ i \delta^{AB} f^{Aijk} \tau^{Ak}$, $f^{1ijk}=0$, $f^{2ijk}=\varepsilon^{ijk}$ 
and  $f^{3ijk}$ is the $SU(3)$ structure tensor. 

All the scalar fields from the second term in Eq.~(\ref{lsmyukawa}) carry, due to their scalar index 
$s=(7,8)$, the weak charge and the hyper charge $\mp \frac{1}{2}$ and $\pm \frac{1}{2}$, 
respectively, as does the Higgs's scalar in the {\it Standard Model}. They carry additional quantum
numbers in adjoint representations (the family quantum numbers originating in $\widetilde{S^{ab}}$ 
and the quantum numbers $Q,Q',Y'$ (we kindly ask the reader to look for the detailed explanation
in Refs.~\cite{M1315,matterantimatter,scalar})  and determine, after they gain nonzero 
vacuum expectation values, together with the vector gauge fields in loop corrections in all orders 
mass matrices of the two groups of four families. These scalar fields predict that at the LHC will be 
measured not only the fourth family quarks, but also several scalar fields~\cite{GN2014}. 

The evaluation of  masses and mixing matrices of the lower four families~\cite{BBLM} suggests that 
the (stable) fifth family masses, which form the dark matter, should be above a few TeV, while 
evaluations of the breaks of symmetries from the starting one (Eq.~\ref{action}) suggests that these 
masses should be far bellow $10^{10}$ TeV.

Following the history of the fifth family members in the expanding universe up to today and 
estimating also the scattering properties of this fifth family on the ordinary matter~\cite{BM}, 
the evaluated masses of the fifth family quarks,  under the assumption that the lowest mass fifth 
family baryon is the fifth family neutron, are in the interval 
\begin{eqnarray}
\label{all}
{\rm a}\,\, {\rm few} \,{\rm times}\, 10 \,{\rm TeV} < m_5 <  10^5 \,\mathrm {TeV}.
\end{eqnarray} 
In the Refs.~\cite{BM} was predicted that if DAMA/LIBRA~\cite{DAMA,rita} really measures the fifth 
family neutrons, also other direct experiments like CDMS~\cite{CDMS}, CREST~\cite{CRESST} 
and XENON100~\cite{XENON100} should in a few years observe the
dark matter clusters. One of the authors of this paper (S.N.M.B, together with B.M.~\cite{BML}) 
put a lot of efforts to understand, why  DAMA/LIBRA measures the dark matter while the other 
experiments do not (yet?)~\cite{BML,rita}.  In Ref.~\cite{rita} one can read the answers 
to several questions which also the reader might have.

\section{The very heavy (fifth family stable) neutron as a candidate for the dark matter}
\label{calculations}

In this section we study the scattering cross section of a very heavy baryon with the ordinary
nucleon, provided that the ratio of masses $\frac{m_{5}}{m_{1}}$ is large enough 
($\ge 10^{3}$). Since the dark matter constituent with the electromagnetic charge would lead
to a substantial Rutherford scattering, what would not be in agreement with the  observed dark
matter properties, we study the baryon with zero electromagnetic charge, estimating  
limits on ($u_{5}-d_{5}$) quark mass differences under which the neutral baryon $n_5$ 
appears as the lightest baryon (Subsect.~\ref{clusters}).

From the (locally quite approximately) known mass density of the dark matter and from the
freezing out procedure of the fifth family neutrons in the evolution of the universe it is deduced
in Ref.~\cite{BM} (Eq. (7)) that the mass of the very heavy fifth family neutrons must be in 
the interval from a few ten TeV to a few 100 TeV. 

The weak cross section, which appears to be independent of the heavy neutron mass
(Eq.~(\ref{Msigma})), is in agreement with the direct experiments of the dark 
matter~\cite{DAMA}, if  the our very heavy neutrons  (Subsect.~\ref{weak}) have a mass
roughly 100 TeV or larger. 

In Subsect.~\ref{strong} we show within the one gluon exchange approximation that for a
heavy neutron with the mass of  $\ge$ 100 TeV the scattering cross section due to the colour 
interaction is much smaller than the cross section due to the weak interaction.

Let us, before starting with the formal evaluations, try to understand what is happening when
a weak charged colourless cluster of very heavy and consequently strongly bound quarks, with
the electromagnetic charge equal to zero, approaches
nucleus of the ordinary (first family weak charged) baryons. Let their relative velocity be small 
so that the kinetic energy of the lighter cluster with respect to their center of mass motion
(that is almost with respect to the heavier cluster) is small in comparison with the binding energy 
of nucleons in the nucleus. The ordinary nucleus interact with the heavy quark cluster through
the weak and through the colour force. 

The size of any ordinary quark in the nucleus is huge in comparison with our very heavy and 
very tiny fifth family neutron. The very quark, which "hits"our heavy neutron through the weak 
and the colour force, reflects elastically from our very heavy neutron. However, since it is 
strongly bound into the ordinary baryon, while baryon is strongly bound into the nucleus, the 
whole nucleus elastically scatter from our heavy neutron (almost) as from the rigid wall, while
our neutron continues its way with almost unchanged direction and velocity.

It is not difficult to calculate the contribution of the weak force to the scattering amplitude for 
such an event. This is done in Susect.~\ref{weak}. The evaluation of the contribution of the 
colour force needs more effort. In Ref.~\cite{BM} it was assumed that within two orders of 
magnitude this scattering amplitude is proportional to the size of the tiny heavy neutron. In  
scattering of ordinary neutrons on ordinary nucleons the contribution of the nuclear (that is of 
the colour) force is in the low enery regime indeed proportional to the size of the nucleons. 
But in ordinary nucleon the quarks are "dressed" into the large gluon cloud which determine 
the quark mass in the nucleon, while in the heavy neutron case one gluon exchange dominates
in the binding energy.

In the present study we take into account that the gluons can "see" the tiny heavy neutron 
only if the heavy neutron polarizes  during the scattering. We calculate dipole-dipole scattering
only. This is done in Subsect.~\ref{strong}.

 \subsection{Is the very heavy neutron the lightest fifth family baryon?}
 \label{clusters}
 
First we calculate the dominant properties of a three-quark cluster, its binding  energy  and 
size (more details can be found in Ref.~\cite{BM}). For this purpose we assume equal  masses
of all the fifth family quarks and we realize that in the regime that $m_{n_{5}} >$ than a few 
TeV  (table I in  Ref.~\cite{BM}) the colour interaction is coulombic - one gluon exchange dominates - 
 and so is also the weak and the electromagnetic interaction, all determined by one massless 
 weak boson and one photon exchange, respectively. 
 
 For three nonrelativistic particles with attractive coulombic like interaction we solve the equations 
 of motion for the Hamiltonian~\cite{BM}
 \begin{eqnarray}
 \label{H}
 H = 3m_{5} + \sum_i \frac{\vec{p}_i^2}{2m_{5}} - \frac{(\sum_i\vec{p}_i)^2}{6m_{5}} 
- \sum_{i<j} \frac{2}{3}\frac{\alpha_{\mathrm{s}}}{r_{ij}}\,.
\end{eqnarray}

The potential energy of the solution can be parameterized as
\begin{eqnarray}
 \label{V}
V_{\mathrm{s}}= - \frac{2}{3} \alpha_{\mathrm{s}} \epsilon, \quad
\epsilon=\langle\sum_{i<j} \frac{1}{r_{ij}}\rangle = 3\eta \alpha_{\mathrm{s}} m_5,
\end{eqnarray}
where $m_5$ is the average mass of quarks in the fifth family. The parameter $\eta$ for a 
variational solution using Jacobi coordinates and exponential profiles was calculated in 
Ref.~\cite{BM}: $\eta=0.66$. The binding energy is then 
\begin{eqnarray}
 \label{E}
E= \frac{1}{2} \, V_{\mathrm{s}}= -E_{\mathrm{kin}} = - \alpha_{\mathrm{s}}^2 \eta m_5\,.
\end{eqnarray}

The splitting of baryons in the fifth family is caused by the ($u_{5}-d_{5}$) mass difference 
(due to the interaction with the scalar fields carrying the weak, the hyper charge and the family
charges interactions), as well as by the potential energy of the electromagnetic and weak 
interaction. 
Even if we are far above the electroweak phase transition we can still work in the basis using 
Weinberg mixing of $\gamma$ and Z, which is familiar to the low energy hadron physicists. 
We shall therefore use this basis.

Let us split the electro-weak interaction into five contributions: i.  electric, ii. Z-exchange 
Fermi (vector), iii. Z-exchange Gamov-Teller (axial),  iv. W-exchange Fermi (vector) and  
v. W-exchange Gamov-Teller (axial) contributions. We obtain for the mass $M$ of the heavy 
cluster the expression
\begin{eqnarray}
 \label{M}
M=\sum_i m_{5i} + E + \left(V_{\mathrm{EM}}+ V_{\mathrm{Z}}^{\mathrm{F}} 
+ V_{\mathrm{Z}}^{\mathrm{GT}}
+ V_{\mathrm{W}}^{\mathrm{F}}+ V_{\mathrm{W}}^{\mathrm{GT}}\right)\,, 
\end{eqnarray}
where the corresponding five contributions are as follows
\begin{eqnarray}
 \label{Vterms}
V_{\mathrm{EM}}&=& \langle\sum_{i<j}q_{5i}\,q_{5j}\rangle\>\alpha_{\mathrm{EM}}\,\epsilon\,,\nonumber\\
V_{\mathrm{Z}}^{\mathrm{F}}&=& \langle\sum_{i<j}(\frac{t^3_i}{2}-\sin^2\vartheta_{\mathrm{W}}q_{5i})\,
(\frac{t^3_j}{2}-\sin^2\vartheta_{\mathrm{W}}q_{5j})\rangle\> \alpha_{\mathrm{Z}}\,\epsilon\,,\nonumber\\
V_{\mathrm{Z}}^{\mathrm{GT}} &=& \langle\sum_{i<j}\frac{t^3_i \,t^3_j}{4}\vec{\sigma}_i\vec{\sigma}_j)
\rangle\> \alpha_{\mathrm{Z}}\,\epsilon\,,\nonumber\\
V_{\mathrm{W}}^{\mathrm{F}}&=& \langle\sum_{i<j}\frac{t^-_it^+_j+t^+_it^-_j}{8}\rangle\>
\alpha_{\mathrm{W}}\,\epsilon\,,\nonumber\\
V_{\mathrm{W}}^{\mathrm{GT}}&=& 
\langle\sum_{i<j}\frac{t^-_it^+_j+t^+_it^-_j}{8}\vec{\sigma}_i\vec{\sigma}_j\rangle\>
\alpha_{\mathrm{W}}\,\epsilon\,.
\end{eqnarray}
The operators $\vec{t}=\frac{1}{2}\vec{\tau}$ are the isospin operators, $t^\pm=(t_1+ i t_2)$, and 
$\vec{\sigma}$ are the Pauli spin matrices, $q_{5i}$ are the electromagnetic charges of quarks. 

The numerical values for all these five terms are, for a particular choice of the parameters, 
presented in table~\ref{EW}, where we choose for the average quark mass $m_5$ = 100 TeV with the 
corresponding average momentum of each quark $p=\sqrt{2m_5\,E_{\mathrm{kin}}/3} = 5.1$~TeV. At 
this momentum scale, we read the running coupling constants  from Particle Data Group 
diagram~\cite{PDG} as $\alpha_{\mathrm{s}}=\alpha_3=1/13$,
 $\alpha_{\mathrm{W}}=\alpha_2=1/32$ 
and $\alpha_1=1/56$. It then follows that $\sin^2\vartheta_{\mathrm{W}}= $ 
$(1+ \frac{5}{3}\frac{\alpha_\mathrm{W}}{\alpha_1})^{-1}= 0.255\approx 1/4 $, 
$\; \alpha_{\mathrm{EM}}=\alpha_{\mathrm{W}} \sin^2\vartheta_{\mathrm{W}}=1/128 \;$ and
$\alpha_{\mathrm{Z}}=\alpha_{\mathrm{W}}/ \cos^2\vartheta_{\mathrm{W}} = 1/24$. 
\begin{table}[hhh]
\centering
\vspace*{4mm}
\begin{tabular}{|l|l|l|l|l|} \hline 
& $u_{5} u_{5} u_{5}$ & $u_{5} u_{5} d_{5} $ & $u_{5} d_{5} d_{5} $ & $d_{5} d_{5} d_{5} $ \\ \hline
$V_{\mathrm{EM}}/\epsilon\alpha_{\mathrm{EM}}$                   & +4/3 & 0     & -1/3 & +1/3 \\ 
$V_{\mathrm{Z}}^{\mathrm{F}}/\epsilon\alpha_{\mathrm{Z}}$  & +1/48 & -1/48 & 0 & +4/48 \\
$V_{\mathrm{Z}}^{\mathrm{GT}}(\mathrm{n}_{5})/\epsilon\alpha_{\mathrm{Z}}$&---&-15/48&-15/48&---\\
$V_{\mathrm{Z}}^{\mathrm{GT}}(\Delta_{5})/\epsilon\alpha_{\mathrm{Z}}$  &-9/48     &+3/48&+3/48&-9/48\\
$V_{\mathrm{W}}^{\mathrm{F}}/\epsilon\alpha_{\mathrm{W}}$                      &0&+1/4&+1/4&0\\
$V_{\mathrm{W}}^{\mathrm{GT}}(\mathrm{n_{5}})/\epsilon\alpha_{\mathrm{W}}$&---&-30/48&-30/48& --- \\
$V_{\mathrm{W}}^{\mathrm{GT}}(\Delta_{5})/\epsilon\alpha_{\mathrm{W}}$         &0&-1/4&-1/4&0\\
\hline
$V_{\mathrm{EW}}(\mathrm{n_{5}})/\epsilon$ &---& -0.0256 & -0.0273 &---\\
$V_{\mathrm{EW}}(\Delta_{5})/\epsilon$ & +0.0035 & +0.0017 & -0.0000 & -0.0017 \\
\hline
\end{tabular}
\caption{Electro-weak contributions to the fifth family - very heavy- baryon masses for the five
terms in Eq.~(\ref{Vterms}). The values for the parameters are presented in the text. The lowest
two lines are the sum over above contributions. The unnecessary decimal places are there if the 
reader likes to check  the reproducibility of the results, --- means non applicable.
One notices that the vector contributions are the same for neutrons ($n_{5}$) and deltas
 ($\Delta_{5}$) baryons, while the axial contributions differ dramatically.}  
\label{EW}
\end{table}

In this example, the binding energy $E=-0.39$ TeV and the average reciprocal distance 
$\langle 1/r_{ij} \rangle = \epsilon/3 = \eta\alpha_{\mathrm{s}}m_5=5.1 \,\mathrm{TeV} 
= 2.6\cdot10^4\, \mathrm{fm}^{-1}$.

One finds for the limits on ($u_{5}-d_{5}$) mass difference, which make the neutral baryon $n_{5}$
the lightest, the following relations
\begin{eqnarray}
\label{limitseq}
&& (m_{u5}-m_{d5}) < (0.0273-0.0017)\,\epsilon = 0.0�256\,\epsilon \,,\quad  {\rm preventing}\,\; 
(u_{5} d_{5} d_{5})\, \to \,(d_{5} d_{5} d_{5}) \, ,\nonumber\\
&& (m_{u5}-m_{d5}) > (-0.0273+0.0256)\,\epsilon = -0.0017\,\epsilon \,,\quad {\rm preventing}\,\;
(u_{5} d_{5} d_{5})\, \to \,(u_{5} u_{5} d_{5}) \,.
\end{eqnarray}
For our value of $\epsilon=15.24$ TeV the above requirements lead to the relation 
\begin{eqnarray}
\label{limitsnum}
- 0.026 \,\mathrm{TeV}\;< (m_{u5}-m_{d5}) <\; 0.39 \, \mathrm{TeV}\,.
\end{eqnarray}
This limits are narrow compared to the mass scale $m_5=100$ TeV, but they are not so narrow if 
the mass generating mechanism would be of the order of 100 GeV as it is in the case of the 
so far observable familes.

\subsection{The weak  (u$_5$d$_5$d$_5$)  --  (u$_1$d$_1$d$_1$)  cross section}
\label{weak}

It is easy to calculate the weak scattering amplitude for (u$_5$d$_5$d$_5$)  --  (u$_1$d$_1$d$_1$),
since the very heavy neutron ($n_{5}$) is a point
particle compared to the range of the weak interaction and its quark structure is negligible.
Only Z-exchange part matters since there is not enough energy to excite (u$_5$d$_5$d$_5$) into 
(d$_5$d$_5$d$_5$)  or  (u$_5$u$_5$d$_5$) via W-exchange. We consider only the scattering on 
neutron (the "charge" of a proton (almost) happens to cancel). Also, we consider only the Fermi
(vector) matrix element, since it contributes coherently in heavy nuclei, while the Gamov-Teller (axial) 
has many cancellations in spin coupling. In Eq.~(\ref{Msigma}) the matrix element and the 
scattering amplitude are presented.
\begin{eqnarray}
\label{Msigma}
{\cal M} &=& [\frac{1}{2}\, t^3_{1} - \sin^2\,\vartheta_W\, q_{1}]\,\frac{g^2_Z}{m^2_Z}\,
[\frac{1}{2}\, t^3_{ 5} - \sin^2\,\vartheta_W \,q_{5}] = \frac{G_\mathrm{F}}{2\,\sqrt{2}}\,,\nonumber\\
\sigma_\mathrm{n} &=& 2\pi\,|{\cal M}|^2\,\frac{4\pi p_1^2}{(2\pi)^3v^2} 
               = \frac{m_{n_{1}}^2}{\pi}\,|{\cal M}|^2 
               = \frac{G_\mathrm{F}^2 m_{n_{1}}^2}{8\pi} = 1.9\times10^{-13} \mathrm{fm}^2\,,
\end{eqnarray}

Index ${}_{1}$ refers to the first family fermions, ${}_{5}$ to the fifth family ones.
We note that the cross section does not depend on the mass $m_{n_{5}}$ provided it is much larger 
than $m_{n_{1}}$ (the first family nucleon mass). For the scattering amplitude of our $n_{5}$ on a 
target with $A$ nucleons and $Z$ protons we have $\sigma_A $
$= \sigma_\mathrm{n} \,(A-Z)^2A^2 $, since at low enough velocity ($100< v < 300$) km/s of
the heavy neutron, $n_{5}$, scattering is coherent~\cite{BM}.\\

Let us now evaluate the rate at a detector of $_{11}^{23}$Na\, $_{53}^{127}$I per kilogram of 
detector 
\begin{eqnarray}
\label{weakrate}
R_{1kg} &=& \sigma_A \,N_A\,\frac{\rho_{n_{5}}\>  v}{m_{n5}}\,\nonumber\\
R_{1kg} &=& \sigma_{n} \,[(A_{Na}-Z_{Na})^2A_{Na}^2 + (A_I-Z_I)^2A_I^2]\,
           \frac{N_{Av}}{A_{Na}+A_I}\,\frac{\rho_{n5}\>  v}{m_{n_{5}}} = 1.3/\mathrm{day}\,,
\end{eqnarray}
where $Av$ is the Avogadro number, and where $\rho_{n5}=0.3\, \mathrm{GeV \,cm}^{-3},\;
m_{n5}=300\,\mathrm{TeV}$ (for $m_{q_{5}}=100$ TeV), $\;v=230\,\mathrm{km/s}$. More 
details can be found in Ref.~\cite{BM}. (The bound is only approximate since  
$(100<v<300)\,\mathrm{km/s}$.)

This is in agreement with the rate claimed by the DAMA/LIBRA~\cite{DAMA} collaboration:
$\Delta R_{1kg}(\mathrm{DAMA})=0.02/\mathrm{day}$, 
$R_{1kg}(\mathrm{DAMA})\sim (0.1\leftrightarrow 1)/\mathrm{day}$.

Should the DAMA results turn out to be smaller,
or would not be  confirmed,  then
$m_{q_{5}}$ should be even larger, if for the 
velocity of Sun $v=230\,\mathrm{km/s}$ is an acceptable value~\cite{BM}.

\subsection{The colour  heavy meson --  light meson  cross section}
\label{strong}

We estimate the scattering  cross section of (u$_5$d$_5$d$_5$) on (u$_1$d$_1$d$_1$),  
or on any first family nucleon, in one boson exchange approximation. A detailed calculation 
is a demanding few body problem even in the lowest order approximation. We get the answer 
to the question to which extent the assumption that this cross section is within a  factor of 
a few $10$ proportional to the square of the size of the fifth family nucleon ($n_{5}$), as it is 
assumed in Ref.~\cite{BM}, by calculating the simpler problem: the ${\rm meson}_{5}$ - 
${\rm meson}_{1}$ scattering within one boson exchange approximation. In this case we
trust  that the baryon in a quark-diquark approximation resembles a meson.

Let us point out that the evaluation, presented in this section, is very relevant also for
bottomium or future heavy baryons in the ($10-100$) GeV region scattering on the first 
family baryons.

In the equation  below~(\ref{trialfunctions}) we present the trial functions of the light and the
heavy meson, together with relevant quantities, such as the chromomagnetic dipole moment 
$D$ (in our case $D_{5}$  of the heavy meson sitting in the dipole field $G_{1}$ of the light 
meson. We use  $m_{1}$ and $m_{5}$ for the light and heavy quark masses, respectively,
and $q_{1}$, $\bar{q}_{1}$, $q_{i}$, $\bar{q}_{i}, i=(1,2)$ represent quark and antiquark
for either light ($i=1$) or heavy ($i=5$ quarks. The interaction strength is 
$\alpha= \frac{4}{3} \alpha_s\,.$ 
$\vec{r}$ and $\vec{R}$ are relative coordinates of light and heavy mesons.The parameters 
$b$ and $B$ are the widths of the hydrogen-like (actually positronium-like) wave functions for
the  light or heavy quark-antiquark pair. In the virtual excited state of the light meson, the 
width $f$ is determined by minimizing the energy of the composite light meson -- heavy meson
system. In the virtual excited state of the heavy meson we keep the same width as in the 
ground state since the change due to minimization was negligible. This is expected since the
light system can adapt easily due to small excitation energies while the heavy system is too 
``expensive''.
\begin{align}
\vec{r}&=\vec{r}_{q_1} - \vec{r}_{\bar{q}_{1}},\quad b=1/({\half m_{1}})\alpha
&\vec{R}&=\vec{R_{q_{5}}} - \vec{R}_{\bar{q}_{5}},\quad B=1/({\half m_{5}})\alpha << b \nonumber\\
\psi_0&=(2/\sqrt{4\pi b^3})\,\exp(-r/b)
&\Psi_0&=(2/\sqrt{4\pi B^3})\,\exp(-R/B) \nonumber\\
\psi_z&=\frac{2^{-3/2}}{\sqrt{4\pi f^3}}\,(r/f)\cos\vartheta\,\exp(-r/f)
&\Psi_z&=\frac{2^{-3/2}}{\sqrt{4\pi B^3}}\,(R/B)\cos\Theta\,\exp(-R/B) \nonumber\\
\epsilon_0&= - (1/2) ({\half m_{1}})\,\alpha^2
&E_0&= - (1/2)({\half m_{5}})\,\alpha^2  \nonumber\\
\epsilon_{z,kin}&= + (1/8) ({\half m_{1}})\,\alpha^2\,(b/ f)^2
&E_z&= - (1/8)({\half m_{5}})\, \alpha^2 \nonumber\\
G_{z_{1}}&=\langle \psi_z|z/(r/2)^3|\psi_0\rangle = \gamma /\sqrt{fb^3} 
&D_{5}&=\langle \Psi_z|z_{5}|\Psi_0\rangle = \beta B  
\label{trialfunctions}
\end{align}
The numerical factors are denoted as $\gamma = 16\sqrt{2}/3 = 7.542 $
and $\beta = 2^{15/2} /3^5 = 0.745 $.

The meson wave functions carry also a factor corresponding to the colour degree of freedom.
The three single particle colour eigenstates (``red'', ``green'' and ``blue'') are denoted as
${\cal R} = (\frac{1}{2}, \frac{1}{2\sqrt{3}})$, \, ${\cal B} = (-\frac{1}{2}, \frac{1}{2\sqrt{3}})$, \,
${\cal G} = (0, -\frac{1}{\sqrt{3}})$  while the anticolour states are 
${\bar{\cal R}} = (-\frac{1}{2}, -\frac{1}{2\sqrt{3}})$, \, 
${\bar{\cal B}} = (\frac{1}{2},- \frac{1}{2\sqrt{3}})$, \,
${\bar{\cal G}} = (0, \frac{1}{\sqrt{3}})$.

The whole states,  for the spatial and the colour part, are then for the light meson
$$\phi_0=\psi_0\,\,\frac{(|{\cal R}> |\bar{{\cal R}}> 
+ |{\cal G}>|\bar{{\cal G}}>+ |{\cal B}> |\bar{{\cal B}}>)}{\sqrt3},\quad 
\phi_{z3}=\psi_{z}\,\frac{(|{\cal R}> |\bar{{\cal R}}> - |{\cal G}>|\bar{{\cal G}}>)}{\sqrt2}\,,$$
and for the heavy meson
$$\Phi_0=\Psi_0\,\frac{(|{\cal R}> |\bar{{\cal R}}> 
+ |{\cal G}>|\bar{{\cal G}}>+ |{\cal B}> |\bar{{\cal B}}>)}{\sqrt3},\quad 
\Phi_{z3}=\Psi_{z}\,\frac{(|{\cal R}> |\bar{{\cal R}}> - |{\cal G}>|\bar{{\cal G}}>)}{\sqrt2}\,.$$

Only the spatial excitation in the $z$-direction and the excitation corresponding to  the  colour
operator $\lambda_q^3$ are in Eq.~(\ref{trialfunctions}) written explicitly. Other directions and
other operators can be written in a similar way.

We need the colour matrix element
$$\left\langle\, \frac{(|{\cal R}> |\bar{{\cal R}}> - |{\cal G}>|\bar{{\cal G}}>)}{\sqrt2}\right\vert
\frac{\lambda_{q_{1}}^3}{2} - \frac{\lambda_{\bar{q}_{1}}^3}{2}
\left\vert \, \frac{(|{\cal R}> |\bar{{\cal R}}> 
+ |{\cal G}>|\bar{{\cal G}}>+ |{\cal B}> |\bar{{\cal B}}>)}{\sqrt3}\right\rangle
=\sqrt{\frac{2}{3}}$$

For  the colour neutral hadrons, the dominant term in the expansion is  the
effective dipole--dipole, colour-octet -- colour-octet potential
$$\hat{V}_{\mathrm{dipole}} = \alpha_s \, \left(\vec{R}_{q_{5}}\,\>\,
\frac{\overrightarrow{\lambda_{q_{5}}}}{2} 
+ \vec{R}_{\bar{q}_{5}}\,\>\,\frac{ \overrightarrow{\lambda_{\bar{q}_{5}}}}{2}\right)\cdot
\left(\frac{\vec{r}_{q_{1}}}{r_{q_{5}}^3}\, \>\, \frac{\overrightarrow{\lambda_{q_{1}}}}{2} 
+ \frac{\vec{r}_{\bar{q}_{1}}}{r_{\bar{q}_{1}}^3}\> \frac{ \overrightarrow{\lambda_{\bar{q}_{1}}}}{2}\right)\,.$$

There are two scalar products, between $\vec{R}$ and $\vec{r}$
and between $\overrightarrow{\lambda}$ and $\overrightarrow{\lambda}$.
The perturbation term between the unperturbed ground state and the virtual excitation is then
$$V_{z,3}=\alpha_{\mathrm{s}} \langle \Psi_z\psi_z|\{\frac{R_{3}}{2}\}\sqrt{\frac{2}{3}}\;
\{\frac{r_{3}/2}{(r/2)^3}\}\sqrt{\frac{2}{3}}| \Psi_0\psi_0\rangle
= \frac{\alpha_{\mathrm{s}}\,D_{5 r_{3}}\,G_{1 _{3}}}{6}$$
$$V_{r_{1},\omega}=V_{r_{2},\omega}=V_{r_{3},\omega} \equiv V
\quad {\mathrm{equal\, for\, all}}\,\,\omega\,.$$
where the index  $\omega=1,.......8$  corresponds to all colour excitations.

The second order perturbation theory then gives the effective potential between the two 
clusters
\begin{eqnarray}
\label{Veff}
V_{eff} &=& -24 \,\frac{{V}^2}{(E_z-E_0) + \epsilon_{z,kin}}\,.\nonumber
\end{eqnarray}
We neglected $\epsilon_{z,pot}$ and $\epsilon_0$. The factor 24 comes from 3 spatial and
8 colour degrees of freedom.
$$V_{eff} = -\frac{2}{3} \
,\frac{(\alpha_{\mathrm{s}} D_{5 z} G_{1z})^2}{(3/8)({\half m_{5}})(4\alpha_{\mathrm{s}}/3)^2 
+ (1/8) ({\half m_{1}})(4\alpha_{\mathrm{s}}/3)^2(b/f)^2} $$
$$V_{eff} = -\frac{2(\beta\gamma\,B)^2}{fb^3(m_5+(1/3)m_{1}(b/f)^2)} $$
Note that $\alpha_{\mathrm{s}}$ has cancelled. Minimization \\
with respect to $f$ gives 
$f/b=\sqrt{m/3M}<<1$. Finally, we get
$$V_{eff} = -\frac{\sqrt{3}\beta^2\gamma^2}{ b^3}\,(\frac{m_{1}}{m_{5}})^{3/2} \,
\frac{B}{m_{1}} $$
Here we take that the distance between the two clusters is equal to zero, $U=0$. We assume 
$$V_{eff}(U) =V_{eff} (U=0)\,\exp(-2U/b).$$
In Born approximation (with the mass of the lighter cluster  $m_{q_{1}} + m_{\bar{q}_{1}}
 = 2m_{1}$) we get
\begin{equation}
a=\frac{(2m_{1})}{2\pi} \int V_{eff} (U) d^3 U =\sqrt{3}\beta^2\gamma^2 
 (\frac{m_{1}}{m_{5}})^{3/2}\,B \,.
\end{equation}
This is our main result, that is the scattering amplitude in the Born approximation,
which is obviously not just proportional to the size of the meson.
In order to see how large is this value in comparison with the size of the heavy meson, which is
of the order of the width $ B=1/({\half m_{5}})\alpha\approx 4\cdot10^{-5}$ fm
let us now give a numerical example with the choice
$$m_1=300\,\mathrm{ MeV},\quad m_{5}=100 \,\mathrm{ TeV}, \quad m_{1}/m_{5}=3\cdot10^{-6},
\alpha_\mathrm{s}=1/13$$
\begin{equation}
a=\sqrt{3}\beta^2\gamma^2  (\frac{m_{1}}{m_{5}})^{3/2}\,B=1.1\cdot10^{-11}\,\mathrm{fm}\,.
\end{equation}
The cross section corresponding to this scattering amplitude is

\begin{equation}
 \sigma=4\pi a^2 = 1.5\cdot10^{-21}\,\mathrm{fm}^2 \,.
\end{equation}

\section{Conclusion}
\label{conclusions}

In this paper we discuss scattering amplitudes of very heavy stable electromagneticaly 
neutral neutrons ($n_{5}$)  on the first family neutrons ($n_{1}$) within the one boson exchange 
approximation. The fifth family stable 
very heavy neutrons are, namely, predicted by the {\it Spin-Charge-family} theory, proposed 
by one of the authors~\cite{M92,M93,M01,M95,M1315,matterantimatter,scalar,Bled} and are 
candidates for the dark matter.

The {\it Spin-Charge-family} theory is very promising to be the right step beyond the {\it 
Standard Model}: It offers an explanation not only for the assumptions of the {\it Standard 
Model}, with the appearance of families included, but also for other phenomenological facts, 
such as the dark matter and the matter-antimatter asymmetry in our universe. We hope 
that the reader will enjoy this theory and would hopefully like to contribute to its application.

The evaluations  of the scattering amplitudes for the stable heavy (fifth family) neutron is 
presented in Sect.~\ref{calculations}. The amplitudes are calculated in the one boson 
exchange approximation, replacing the heavy and the first family neutrons ($n_{5}$, $n_{1}$) 
with the simpler meson$_{5}$ and meson$_{1}$ to check whether a rough estimation~\cite{BM}, 
that the scattering  amplitude caused by the colour force is proportional to the size of 
$n_{5}$, is an acceptable approximation for very heavy neutrons. It turned out  that the 
scattering amplitude at least in the colour dipole - colour dipole approximation is much smaller than the 
one estimated roughly in Ref.~\cite{BM}.

This paper is written also in purpose to convince the hadron physicists that if  the 
{\it Spin-Charge-Family} theory is the right step beyond the {\it Standard Model} then they 
will have a pleasant time to study properties of the clusters forming dark matter with their 
knowledge from the lower three families. It is interesting to notice how much does the 
"nuclear" physics change when the constituents are very heavy. We demonstrate in 
Sect.~\ref{calculations} that for the mass of the fifth family neutron  ($n_{5}$) equal or 
greater than $100$ TeV, the weak force is stronger than the 
"fifth family-first family" nuclear force. 

We also offer the present calculations as an introduction to study of heavy hadron -- light 
hadron scattering.

 Although either meson-meson or the one boson exchange approximations, used in our 
calculation, might not be accurately enough 
yet the extremely small ratio between the weak and the "fifth family-first family" nuclear 
force  in the case that $m_{n_{5}}\approx $  100 TeV or larger tells that the nuclear force 
among very heavy objects is very very weak.


\begin{thebibliography}{99}

\bibitem{M92} N. S. Manko\v{c} Bor\v{s}tnik, {\it Phys. Lett.} {\bf B 292}, 25 (1992).
\bibitem{M93} N. S. Manko\v{c} Bor\v{s}tnik, {\it J. Math. Phys.} {\bf 34}, 3731 (1993).
\bibitem{M01} N. S. Manko\v{c} Bor\v{s}tnik, {\it Int. J. Theor. Phys.} {\bf 40}, 315 (2001).
\bibitem{M95} N. S. Manko\v{c} Bor\v{s}tnik, {\it Modern Phys. Lett.} {\bf A 10}, 587 (1995) 587.
\bibitem{M1315} N. S. Manko\v{c} Bor\v{s}tnik, {\it J. of Modern Phys.} {\bf 4}, 823 
(2013) [arxiv:1312.1542].
\bibitem{matterantimatter} N. S. Manko\v{c} Bor\v{s}tnik, {\it Phys. Rev.} {\bf D 91}, 
065004 (2015) [arxiv:1409.7791].
%
\bibitem{scalar}  N. S. Manko\v{c} Bor\v{s}tnik, 
"The {\it spin-charge-family} theory explains why  the scalar Higgs carries the weak charge 
$\pm \frac{1}{2}$ and the hyper charge $ \mp \frac{1}{2}$", [arxiv:1312.1542], Proceedings 
to the $17^th$ Workshop "What comes beyond the standard models", Bled, 20-28 of July, 2014, 
Ed. N.S. Manko\v c Bor\v stnik, H.B. Nielsen, D. Lukman, DMFA  Zalo\v zni\v stvo, 
Ljubljana December 2014, p.163-82 [ arXiv:1502.06786v1] [http://arxiv.org/abs/1409.4981].
\bibitem{AM} A. Bor\v{s}tnik, N. S. Manko\v{c} Bor\v{s}tnik, {\it Phys. Rev. D} {\bf 74}, 073013 (2006),
 hep-ph/0512062; in {\it Proceedings to the Euroconference on Symmetries Beyond the Standard Model}, 
Portoro\v{z}, July 12-17, 2003, hep-ph/0401043, hep-ph/0401055, hep-ph/0301029.
\bibitem{BM} G. Bregar and N. S. Manko\v{c} Bor\v{s}tnik, {\it Phys. Rev. D} {\bf 80},
 083534 (2009) [arXiv:1412.5866].
\bibitem{BBLM} G. Bregar, M. Breskvar, D. Lukman, N.S. Manko\v{c} Bor\v{s}tnik, 
{\it New J. of Phys.} {\bf 10}, 093002 (2008). 
\bibitem{GN2014} G. Bregar, N.S. Manko\v c Bor\v stnik, "The new experimental data for 
the quarks mixing matrix are in better agreement with the {\it spin-charge-family} theory 
predictions", Proceedings to the $17^{th}$ Workshop "What comes beyond the standard models", 
Bled, 20-28 of July, 2014, Ed. N.S. Manko\v c Bor\v stnik, H.B. Nielsen, D. Lukman, DMFA 
Zalo\v zni\v stvo, Ljubljana December 2014, p.20-45 [arXiv:1502.06786v1] 
[arxiv:1412.5866].
\bibitem{Bled} N.~S.~Manko\v{c} Bor\v{s}tnik, "Mass matrices of twice four families of
quarks and lepton, scalars and gauge fields as predicted by the spin-charge-family theory",
Proceedings to the $13^{th}$ Workshop 
"What comes beyond the standard models", Bled, 12-22 of July, 2010, 
Ed. N. S. Manko\v c Bor\v stnik, H. B. Nielsen, D. Lukman, DMFA 
Zalo\v zni\v stvo, Ljubljana December 2010, p.105--129.
\bibitem{DAMA} R. Bernabei et al., Int. J. Mod. Phys. D {\bf 13} (2004) 2127-2160;
Eur. Phys. J. C {\bf 67} (2010) 39, [arXiv:1412.6524]. 
\bibitem{rita} R. Bernabei, "Dark matter particles in the galactic halo",  Proceedings to 
the $17^th$ Workshop "What comes beyond the standard models", 
Bled, 20-28 of July, 2014, Ed. N.S. Manko\v c Bor\v stnik, H.B. Nielsen, D. Lukman, DMFA 
Zalo\v zni\v stvo, Ljubljana December 2014, p.10-19 [arXiv:1502.06786v1], 
[arxiv:1412.6524].
\bibitem{CDMS} Z. Ahmed et al., {\it  Phys. Rev. Lett} {\bf 102}, 011301 (2009).
\bibitem{CRESST}
G. Angloher at al., {\it Eur. Phys. J} {\bf 72}, 197 (2012) 19.; 
\bibitem{XENON100}
E. Aprile at al {\it Phys. Rev. Lett.} {\bf 111}, 021301 (2013).
\bibitem{BML} G. Bregar, R. F. Lang,  N. S. Manko\v{c} Bor\v{s}tnik, "The prediction of the 
spin-charge-family theory that the fifth family neutrons constitute the dark Matter in 
with the XENON100 experiment, Proceedings to the $13^{th}$ Workshop "What comes beyond the 
standard models", Bled, 12-22 of July, 2010, Ed. N. S. Manko\v c Bor\v stnik, H.B. Nielsen, 
D. Lukman, DMFA Zalo\v zni\v stvo, Ljubljana December 2010, p.161-165 [arXiv:1012.0224].
\bibitem{K} M.Y. Khlopov, {\it Mod. Phys. Lett.}  {\bf A 26}, 2813 (2011) [arXiv:1111.2838 [astro-ph.CO]].
\bibitem{DATA?} A. Ivanov [arxiv:1308.3084v3]. 
\bibitem{Vysotsky}  A.N.Rozanov, M.I. Vysotsky, [arxiv:1012.1483, arxiv:1312.0474]. 
\bibitem{MN1012} N. S. Manko\v{c} Bor\v{s}tnik, M. Rosina, "Are superheavy quark clusters 
viable candidates for the dark matter", Proceedings to the $13^{th}$ Workshop 
"What comes beyond the standard models", Bled, 12-22 of July, 2010, 
Ed. N.S. Manko\v c Bor\v stnik, H.B. Nielsen, D. Lukman, DMFA 
Zalo\v zni\v stvo, Ljubljana December 2010, p.203--210 [arXiv:1012.0224].
\bibitem{Bled2010} N. S. Manko\v{c} Bor\v{s}tnik and M. Rosina,
{\it Bled Workshops in Physics} {\bf 11}  No. 1, 64 (2010) ;
also http://www-f1.ijs.si/BledPub/.
\bibitem{Bled2012} N. S. Manko\v{c} Bor\v{s}tnik and M. Rosina,
{\it Bled Workshops in Physics} {\bf 13}  No. 1, 78 (2012) ;
also http://www-f1.ijs.si/BledPub/. 
\bibitem{PDG} K. Nakamura et al. (Particle Data Group), {\it J. Phys.} {\bf G 37} 075021 (2010).


\end{thebibliography}
\end{document}